\documentclass[12pt]{article}
\usepackage{amsxtra,amssymb,amsthm,amsmath,latexsym}

\theoremstyle{plain}

\def\oH{{\overset{\circ}{H}}}
\def\oH1{{\overset{\circ}{H}\kern-.02in{}^1}}

\def\bee{\begin{equation*}}
\def\eee{\end{equation*}}
\def\be{\begin{equation}}
\def\ee{\end{equation}}

\begin{document}

\title{ Scattering of electromagnetic waves by
small impedance particles of an arbitrary shape}

\author{Alexander G. Ramm\\
 Department  of Mathematics, Kansas State University, \\
 Manhattan, KS 66506, USA\\
ramm@math.ksu.edu\\
http://www.math.ksu.edu/\,$\widetilde{\ }$\,ramm}

\date{}
\maketitle\thispagestyle{empty}

\begin{abstract}

An explicit formula is derived for the  electromagnetic (EM) field scattered by one small impedance
particle $D$ of an arbitrary shape. If $a$ is the characteristic size of the particle, $\lambda$ is the wavelength,
 $a<<\lambda$ and $\zeta$ is the boundary impedance of $D$, $[N,[E,N]]=\zeta [N,H]$ on $S$, where $S$ is the surface
 of the particle, $N$ is the unit outer normal to $S$, and $E$, $H$ is the EM field, then the scattered
 field is $E_{sc}=[\nabla g(x,x_1), Q]$. Here $g(x,y)=\frac{e^{ik|x-y|}}{4\pi |x-y|}$, $k$ is the wave number,
  $x_1\in D$ is an arbitrary point, and
 $Q=-\frac{\zeta |S|}{i\omega \mu}\tau \nabla \times E_0$, where $E_0$ is the incident field, $|S|$ is the area of $S$,
 $\omega$ is the frequency, $\mu$ is the magnetic permeability of the space exterior to $D$, and $\tau$ is a tensor which is calculated explicitly.
  The scattered field is $O(|\zeta| a^2)>> O(a^3)$
 as $a\to 0$ when $\lambda$ is fixed and $\zeta$ does not depend on $a$. Thus, $|E_{sc}|$ is much larger than the classical value $O(a^3)$ for the field  scattered by a small particle.
 It is proved that the effective field in the medium, in which many small particles are embedded, has a limit as $a\to 0$
 and the number $M=M(a)$ of the particles tends to $\infty$ at a suitable rate. This
 limit solves a linear integral equation. The refraction coefficient of the limiting medium is calculated analytically.
 This yields a recipe for creating materials with a desired refraction coefficient.

\end{abstract}

\section{Introduction}\label{S:1}
Let $D\subset \mathbb{R}^3$ be a bounded domain with a connected smooth boundary $S$,
$D':= \mathbb{R}^3\setminus D$, $k^2=const>0$ is the wave number, $\omega>0$ is frequency,
 the boundary impedance  $\zeta=const$, Re $\zeta\ge 0$, $\epsilon>0$ and $\mu>0$ are dielectric and magnetic
constants, $\epsilon'=\epsilon +i\frac{\sigma}{\omega}$, $\sigma=const\ge 0$ is the conductivity of $D'$, $x\in D'$,
$r=|x|$, $N$ is the unit normal to $S$ pointing into $D'$.

Let us assume that the electrical field $E=E_0+e$, where $E_0$ is the incident field and $e$ is the scattered
field. Then $e$ solves the problem:
\be\label{e1} \nabla \times e=i\omega \mu h,\qquad \nabla \times h=-i\omega \epsilon'  h \qquad \text {in}\, D', \ee
\be\label{e2} r(e_r-i k e)=o(1), \qquad r\to \infty, \ee
\be\label{e3} [N,[e,N]]-\frac {\zeta}{i\omega \mu}[N, \nabla \times e]=-f    \text {\,\, on\,\,} S.\ee
Here $f=[N,[E_0,N]]-\frac {\zeta}{i\omega \mu}[N,\nabla \times E_0]$ is a given smooth tangential field on $S$,
$H=\frac {\nabla \times E}{i\omega \mu}$, $h=\frac {\nabla \times e}{i\omega \mu}$, $\epsilon'$ and $\mu$ are
dielectric and magnetic constants of the medium $D'$,
$[A,B]=A\times B$ is the cross product of two vectors,
$A\cdot B$ is their scalar product. Problem (1)-(3)
 is the scattering problem for electromagnetic (EM) waves
by an impedance body $D$ of an arbitrary shape. This problem has been discussed in \cite{R635}, where
 the uniqueness of its solution has been proved. Explicit formula for the plane
EM wave scattered by a small impedance body ($ka\ll 1$, $a$ is the characteristic size of this body)
of an arbitrary shape is derived in \cite{R635}. There one can also find a solution to many-body
scattering problem in the case of small impedance particles (bodies) of an arbitrary shape.

 A few  historical remarks are in order. The theory of wave scattering by small bodies was
originated by Rayleigh in 1871. He understood that the main term in the scattered field is the dipole radiation.
How to calculate this radiation, in other words, how to calculate the induced dipole moment
on a small body of an arbitrary shape, Rayleigh did not show. This was done nearly 100 years later
by the author, see \cite {R476}. In 1908 G.Mie published a method for solving scattering problems for well conducting spherical particles
using separation of variables in the spherical coordinates. His method works also for spherical impedance particles,
but does not work for particles of an arbitrary shape for which separation of variables cannot be used.
Smallness of the particle is not required by the Mie's method.

 There were no analytical methods for solving EM wave
scattering problems for small bodies of an arbitrary shape. No explicit formulas for the scattered fields
were obtained by other authors for small bodies of an arbitrary shape. Such methods for acoustic and EM waves were
developed in the monograph \cite{R635}.

 Let us highlight the novel points in our theory: 
 
 a) For one small impedance particle of an arbitrary shape an analytic formula for the scattered
 field is derived; this formula is asymptotically exact as $a\to 0$; the scattering
 amplitude is $O(\zeta |S|)$. Assuming $|S|=O(a^2)$ and $\zeta=O(a^{-\kappa})$, $\kappa \in [0,1)$
 is a constant, one obtains that the scattering amplitude is $O(a^{2-\kappa})>>O(a^3)$,
 where $O(a^3)$ is the value of the scattering amplitude in the classical theory. This is
 a new physical phenomenon. If the particle is perfectly conducting, the corresponding scattering
 amplitude is $O(a^3)$ for small $a$.
 
 b) In the case of many small impedance particles a method for solving the EM wave scattering
 problem is developed (see formulas (10)-(12)) and the limiting integral equation (see formula (13))
is derived  for the limiting effective field in the medium where very many small impedance particles are embedded.

c) These results are used to formulate a method for creating materials with a desired
refraction coefficient, see Section 5.

 These results do not intersect with the results published by other authors.

\section{Formula for the solution of the EM wave scattering by one small body}\label{S:2}

Our {\em first result} is the following explicit formula for the scattered field:
\be\label{e4} e=[\nabla g(x,x_1), Q],   \qquad g(x,y):=\frac{e^{ik|x-y|}}{4\pi |x-y|},
\ee
which is asymptotically, as $a\to 0$, exact. The quantity $Q$ in formula (4) is given by the following formula:
\be\label{e5} Q=-\frac{\zeta |S|}{i\omega \mu}\tau \nabla \times E_0,
\ee
where $E_0$ is the incident field, the tensor $\tau$ is defined as follows:
\be\label{e6} \tau=I-b, \qquad b=(b_{jm})=\frac 1{|S|}\int_S N_j(s)N_m(s)ds,
\ee
and $ds$ is an element of the surface area. In formula (5) the terms of the higher order
of smallness, as $a\to 0$, are neglected.

 When we write $a\to 0$, it means that $\lambda$
is fixed:  the physical meaning has the ratio $a/\lambda$.

Formulas (4)-(6) are proved in \cite{R635}. In their derivation a new representation of the scattered
field $e$ is used:
$$ e=\nabla \times \int_S g(x,t)J(t)dt, \qquad \int_SJ(t)dt:=Q,$$
where $J$ is a tangential to $S$ field. One can find $e$ of this form
if $k^2$ is not a Dirichlet eigenvalue of the Laplacian in $D$.
If $k^2$ is an eigenvalue of the Laplacian in $D$, then one can use
$g_\rho(x,y)$ in place of $g(x,y)$. Here $g_\rho(x,y)$ is the Green's function
of the Helmholtz operator in the exterior of  $B(0,\rho)$, the ball centered at the origin and of radius $\rho$.
The  $g_\rho(x,y)$ satisfies the Dirichlet condition on the boundary of  $B(0,\rho)$ and the radiation condition at infinity.
 The small number $\rho>0$ is chosen
so that $k^2$ is not a   Dirichlet eigenvalue of the Laplacian in $D\setminus B(0,\rho)$.
 Such a ball
can always be found and $\rho>0$ can be chosen as small as one wishes.

Let us discuss formulas (4)-(6). The choice of $x_1\in D$ in formula (4) is not important since $a$
is small. If $D$ is centrally symmetric body, then one can take as $x_1$ its symmetry center.
Formula (5) shows that the scattering amplitude $A$,  $A=O(|\zeta||S|)=O(a^2)$ if $\zeta$
does not depend on $a$. The classical dependence of the scattering amplitude on $a$ is
$O(a^3)$, which is much smaller than $O(a^2)$ as $a\to 0$.

This fact might be of physical interest in applications.

Formula (6) gives explicitly the  dependence of the scattered field on the shape of the small body.

The main physical (and mathematical) idea, used in  \cite{R635} for the derivation
of the above results is the reduction of the scattering problem for a small particle to finding
just one quantity $Q$ rather than a boundary function.
In the next Section the many-body scattering problem is discussed.

\section{EM wave scattering by many small impedance particles}\label{S:3}
Let us formulate our basic results for EM wave scattering by many
small impedance particles $D_m$, $1\le m \le M$, distributed in a bounded domain $\Omega$. Let $x_m\in D_m$
be  some points.
Assume  that the number $\mathcal{N}(\Delta)$ of small bodies (or points
$x_m$) in any sub-domain $\Delta\subset \Omega$ is given by the formula
\be\label{e7} \mathcal{N}(\Delta)= \frac{1}{a^{2 - \kappa}}\int_\Delta N(x)dx(1 + o(1)),\quad a
\to 0,
\ee
 where $N(x) \geq 0$ is a continuous in $\Omega$ function,
 $\kappa\in [0,1)$
is a parameter, and the boundary impedance of the $m-$th small particle is $\zeta_m:=h(x_m)/a^{\kappa}$,
where $h(x)$ is a continuous function in $\Omega$, such that Re$h\ge 0$, and $x_m\in D_m$ is an arbitrary point.

The choice $\zeta=h(x)a^{-\kappa}$ is not dictated by  physical laws. As we mentioned earlier,
the choice of $\zeta$  is made by the experimentalist as he wishes.

{\em The only  physical restriction on the
boundary impedance is the relation Re$\zeta \ge 0$, that is, Re$h(x)\ge 0$.}

 The restriction for the parameter $\kappa\in [0,1)$ is of technical nature and is not related to a physical law.

{\em The functions $N(x)$, $h(x)$ and the parameter $\kappa$  can be chosen by the
experimentalist as he/she wants.}

Our main physical assumption is:
\be\label{e8}  a<<d<<\lambda,
\ee
where $d$ is the minimal distance between neighboring small particles.

Our {\em second result} is the following formula for the solution of many-body EM wave scattering problem:
\be\label{e9} E(x)=E_0(x)+\sum_{m=1}^M [\nabla g(x,x_m), Q_m], \qquad a\to 0,
\ee
where $Q_m$ are defined by the formula:
\be\label{e10}  Q_m=-\frac{\zeta_m |S|}{i\omega \mu}\tau \nabla \times E_{em}.
\ee
Here we assumed for simplicity that the particles have the same shape, so
tensor $\tau$ does not depend on $m$, and $E_{em}$ is the effective field acting on the $m-$th particle.
This field is defined by the formula:
\be\label{e11} E_{em}=E_0(x_m)+\sum_{j=1, j\neq m}^M [\nabla g(x_m,x_j), Q_j].
\ee
Formulas (10)-(11) lead to the linear algebraic system for finding the unknown quantities $E_m:=E_{em}$:
\be\label{e12} E_{m}=E_0(x_m)-\frac{c_S}{i\omega \mu}\sum_{j=1, j\neq m}^M [\nabla g(x_m,x_j),\tau \nabla \times E_{j}]h_ja^{2-\kappa},\quad 1\le m\le M,
\ee
where $c_S>0$ is the constant in the formula $|S|=c_S a^2$.

Our {\em third result} is the following integral equation for the limiting effective field in $\Omega$:
\be\label{e13} E(x)=E_0(x)-\frac{c_S}{i\omega \mu}\nabla \times \int_\Omega g(x,y),\tau \nabla \times E(y)h(y)N(y)dy,
\ee
where $N(y)$ is defined in formula (7) and the limit is taken as $a\to 0$. The existence of this limit is
proved in \cite{R635}.

Equation (13) is equivalent to the following {\em local} differential equation:
 \be\label{e14} \nabla \times \nabla \times E(x)=k^2 E(x)-\frac {c_S}{i\omega \mu}\nabla \times \Big(h(x)N(x)\tau \nabla \times E(x)\Big),
\ee
as one can check by applying the operator $ \nabla \times \nabla \times$ to equation (13), see the details in \cite{R635}.

Let us assume that $\tau $ is proportional to a diagonal matrix $I$. This happens, for example, if the particles are balls
of radius $a$, in which case $\tau=\frac {2}{ 3} I$, as one can easily verify.

 Then  equation (14)  takes the form:
 \be\label{e15} \nabla \times \nabla \times E(x)=\frac{k^2 E(x)}{1+\frac {2c_S}{3i\omega \mu}h(x)N(x)}-
 \frac {2c_S}{3i\omega \mu}\cdot\frac{[\nabla \Big(h(x)N(x)\Big), \nabla \times E(x)]}{1+\frac {2c_S}{3i\omega \mu}h(x)N(x)}.
\ee

\section{Physical interpretation of formula (15)}\label{S:4}
To interpret physically formula (15), consider the Maxwell's equations:
 \be\label{e16} \nabla \times E(x)=i\omega \mu H, \qquad  \nabla \times H(x)=-i\omega\epsilon' H,
\ee
where $\mu=\mu (x)$, apply the operator $\nabla \times$ to the first equation and then use the second one. This yields:
 \be\label{e17} \nabla \times\nabla \times E(x)=K^2 E(x)+ [\frac{\nabla \mu (x)}{\mu(x)}, \nabla \times E(x)],
\ee
where $K^2=\omega^2 \mu(x)\epsilon'=k^2 \frac {\mu(x)\epsilon'}{\mu_0 \epsilon_0}$, $k^2=\omega^2\epsilon_0 \mu_0$,
 and $\mu_0,\epsilon_0$ are the parameters of the free space. Let $n(x):=\frac{\mu(x)\epsilon'}{\mu_0 \epsilon_0}$.

 Comparing formulas (15) and (17) one concludes that the limiting medium,  obtained by the embedding of many
small impedance particles, has {\em the new refraction coefficient}:
\be\label{e18} n(x)= \frac {n_0(x)}{\Big(1+\frac {2c_S}{3i\omega \mu}h(x)N(x)\Big)^{1/2}}, \qquad n_0(x):=\Big(\frac{\epsilon' \mu}{\epsilon_0\mu_0}\Big)^{1/2},
\ee
and {\em the new magnetic permeability}:
 \be\label{e19} \mu(x)=\Big(1+\frac {2c_S}{3i\omega \mu}h(x)N(x)\Big)^{-1}.
\ee
On formula (18) our recipe  for creating materials with a desired refraction coefficient is based.
This recipe is discussed in the next Section.
\section{A recipe for creating materials with a desired refraction coefficient}\label{S:5}
Let us rewrite formula (18) as
\be\label{e20} n(x)= \frac {n_0(x)}{\Big(1-ic_1h(x)N(x)\Big)^{1/2}},
\ee
where $c_1:=\frac {2c_S}{3\omega \mu}>0$ and $h=h_1+ih_2$, $h_1\ge 0$. The functions
$h(x)$ and $N(x)\ge 0$ are at the disposal of the experimentalist, as was mentioned earlier.
By choosing these functions properly, one can get any desired refraction coefficient which has the property Im $n(x)\ge 0$.
One rewrites formula (20) as follows:
\be\label{e21} n(x)=n_0(x) \Big(1-ic_1h(x)N(x)\Big)^{-\frac 1 2}=n_0(   x)\Big(1+c_1h_2(x)N(x)-ic_1h_1(x)N(x)\Big)^{-\frac 1 2}.
\ee
Let $z$ be a complex number. Define $z^{1/2}=|z|^{1/2}e^{i\phi/2}$, where $\phi$ is the argument of $z$,
$0\le \phi<2\pi$. By choosing  $h_2<0$ so that $ 1+c_1h_2(x)N(x)$ is small and choosing $h_1\ge 0$ suitably,
one can make $|n(x)|$ to be any desired non-negative function. The argument $\phi$ of the expression
$$1+c_1h_2(x)N(x)-ic_1h_1(x)N(x)$$
can be made arbitrary by choosing $h_1\ge 0$ and $h_2\in (-\infty, \infty)$
suitably.

 For example, assume that $n_0>0$, $h_1>0$ is small and $1+c_1h_2(x)N(x)>0$. Then the $\phi=2\pi-2\delta$,
where $\delta$ is arbitrarily small,  $\phi/2= \pi-\delta$ and $n(x)=|n(x)|e^{-i(\pi-\delta)}$.
Thus, Re $n(x)<0$, Im $n(x)\ge 0$, and  Im $n(x)=|n(x)|\sin \delta$ can be made
as small as one wishes if $\delta$ is sufficiently small. Therefore, one gets a material with negative refraction and negligible losses.

Similarly, using formula (19), one can change magnetic permeability in a desired direction by embedding in a given medium many small impedance particles with the suitable boundary impedances.

\end{document}